\documentstyle[12pt,epsf]{article}    
\newcommand{\ra}{\rangle}
\newcommand{\la}{\langle}

\newcommand{\be}{\begin{equation}}
\newcommand{\ee}{\end{equation}}
\newcommand{\ben}{\begin{eqnarray}\displaystyle}
\newcommand{\een}{\end{eqnarray}}

\newcommand{\ba}{\begin{array}}
\newcommand{\ea}{\end{array}}
\newcommand{\bea}{\begin{eqnarray}}
\newcommand{\eea}{\end{eqnarray}}
\newcommand{\beas}{\begin{eqnarray*}}
\newcommand{\eeas}{\end{eqnarray*}}

\def\N{{\cal N}}

\begin{document}
\begin{titlepage}

\begin{flushright}
PUPT-1967\\
SU-ITP 00-26\\
hep-th/0011044
\end{flushright}

\vskip 2.5cm

\begin{center} {\Large \bf

A Note on Dilaton Absorption and

\vspace{4mm}

Near-Infrared D3 Brane Holography} 
 
\end{center}

\vspace{1ex}

\begin{center}
{\large
Leonardo Rastelli$^a$ and Mark Van Raamsdonk$^b$}

\vspace{5mm}
${}^a${ \sl Department of Physics} \\
{ \sl Joseph Henry Laboratories} \\
{ \sl Princeton University} \\
{ \sl Princeton, New Jersey 08544, U.S.A.} \\
{ \tt rastelli@feynman.princeton.edu}

\vspace{5mm}
${}^b${ \sl Department of Physics} \\
{ \sl Stanford University} \\
{ \sl Stanford, CA 94305 U.S.A.} \\
{ \tt mav@itp.stanford.edu}

\end{center}

\vspace{2.5ex}
\medskip
\centerline {\bf Abstract}

\bigskip

We consider the first subleading terms in the low-energy cross section
for the absorption of dilaton partial waves by D3-branes. We
demonstrate that these corrections, computed previously via
supergravity, can be reproduced exactly in a worldvolume calculation
using a deformation of $\N=4$ SYM theory by a dimension eight 
chiral operator. The calculation does not depend on how the theory is 
regularized. This result provides another hint
that holographic duality between the D3-brane
worldvolume theory and the corresponding supergravity solution may be valid
beyond the near horizon limit.

\bigskip

\end{titlepage}

\newpage

\section{Introduction}

The study of particle absorption by D-branes \cite{klebanov,
gubserklebanov, gubsertseytlinklebanov} 
provided one of the early 
hints of an exact correspondence between the gauge theories living on
branes and gravitational physics in the corresponding supergravity p-brane
backgrounds. 
The correspondence is suggested by the existence of two
different pictures of the absorption process. 

In the first, the absorption
is viewed semiclassically as a wave propagating in the appropriate p-brane
supergravity solution and being ``absorbed'' at the horizon. The cross
section is determined by solving the wave equation for the particle of
interest in this geometry with the boundary condition that the wave is
purely ingoing at the horizon. 

The second (``worldvolume'') picture treats
the incident particle as an excitation in the field theory on the
brane  which is absorbed by
decaying quantum mechanically into two or more particles confined to the
brane world volume. From this point of view, the cross section may be most
easily determined by computing the two point function of the
worldvolume operator which couples to the bulk supergravity
particle of interest.

Probably the simplest example to consider is the absorption of a
minimally coupled scalar by a stack of $N$ D3-branes \cite{klebanov, 
gubserklebanov, alecsamir}. 
The supergravity
description is reliable if the energy $\omega$
 of the incident wave is small and the
curvature radius $R$ is large in string units, $\omega \sqrt{\alpha'} \ll 1$,
$R /\sqrt{\alpha'} \gg 1$.  
As already noticed in \cite{klebanov}, the dimensionless combination
$\omega R$ can be kept fixed and arbitrary within the supergravity regime.
In the low-energy limit $\omega R \ll 1$  the interaction of the
incoming partial wave effectively takes place in the near-horizon AdS region.
In the field theory picture, this is the limit in which the brane
degrees of freedom decouple from the bulk and 
are controlled by the superconformal
$\N=4$ SYM theory: this is part of the motivation for the
standard AdS/CFT correspondence \cite{correspondence} between
the $\N=4$ theory and supergravity/string theory on $AdS_5 \times S^5$.

Exact agreement has 
been demonstrated between the supergravity and worldvolume calculations
of the leading low-energy 
absorption cross sections for all partial waves
of the dilaton field \cite{mark}. 
For the $l$-th partial wave, the result of both
calculations is \cite{gubsertseytlinklebanov, mark}
\be
\sigma^l_{wv} = \sigma^l_{sg} = { \pi^4 \over 24}{(l+3)(l+1) \over 
[(l+1)!]^4 2^{4l}} R^{4l+8}  \omega^{4l+3}
\ee
where  $R^4 = 4 \pi g N \alpha'^2$. 
Since the field theory
computation is performed in the free field theory approximation,
while supergravity is valid at strong `t Hooft coupling $g N$,
this exact agreement can only be explained
by assuming a  non-renormalization theorem 
\cite{gubserklebanov}
for the 2-point functions of the $\N =4$ SYM operators dual to the dilaton
partial waves\footnote{Non-renormalization theorems
for the $\N =4$ theory
are now widely believe to hold for all two and three point functions
of chiral operators \cite{nonrenorm, kostas}, 
and even for some special (``extremal'' and ``next-to-extremal'')
 $n$-point functions \cite{DFMMR, extremal}.
See also \cite{also}.}.

Given the precise agreement between the leading order cross sections in the two
pictures, it is interesting to ask what happens away from the low energy
limit. On the supergravity side, the absorption cross section has been
determined exactly by Gubser and Hashimoto  \cite{GH}
as a perturbative expansion to all orders in 
$\omega R$ with the result
\bea
\sigma^l &=& \sigma^l_0 \; (1 + \sum_{n=1}^\infty \sum_{k=0}^n 
 b^l_{n,k}(\omega R)^{4n}(\log(\omega \bar{R}))^k) \nonumber \\ 
&=& \sigma^l_0 \; (1 + b^l_{1,1} (\omega R)^4
\log(\omega \bar{R}) + b^l_{1,0} (\omega R)^4 + \cdots) 
\label{exact}
\eea 
where $\bar{R} = \gamma R /2$ and $b^l_{n,k}$ is a series of numerical
coefficients given implicitly in \cite{GH}\footnote{
For generalizations of this result
see e.g.  \cite{extensions}.}. 
In this paper, we ask whether it
is possible to reproduce any of the correction terms (i.e. determine the
coefficients $b^l_{n,k}$) through a dual calculation in the worldvolume
theory. 

Away from the low-energy limit, the worldvolume theory of the branes
is no longer described simply by $\N=4$ SYM theory. At weak string
coupling, the corrections (for a slowly varying field strength) are
given by the incompletely known non-abelian Born-Infeld action,
\be
\label{BI}
S = {1 \over 2 \pi g} \int d^4x \, \left\{ -{1 \over 4} {\rm Tr}(F^2) + {1 \over 8} (2
\pi \alpha')^2 {\rm STr} (F^4 - {1 \over 4} (F^2)^2) + \dots \right\}  \\ 
\ee
where we have written only the gauge field terms. It is important to
note that the dimension eight term in the Lagrangian,
\be
\label{BIop}
{\cal O}^{BI}_8 = { \pi \over 4 g} {\rm STr} (F^4 - {1 \over 4}
(F^2)^2 + \dots )
\ee
lies in a short 
multiplet of the $\N=4$ supersymmetry algebra\footnote{In fact, it is
the only single-trace non-renormalizable operator which is in a short 
multiplet, is a Lorentz scalar and
preserves the SU(4) R-symmetry. The same operator
arises in the low-energy effective action of
$\N=4$ SYM on the Coulomb branch,
where it leads to logarithmic corrections
to absorption by split D3 branes.
The leading correction 
to dilaton s-wave absorption by a double-centered
D3 brane geometry
has been matched exactly with field theory in \cite{costa}.}, 
so that its dimension is 
protected for any value of the coupling. 
On the other hand, the
higher order corrections in (\ref{BI}) are operators in long multiplets 
(dual to string states) and
acquire large anomalous dimensions in the limit
of strong 't Hooft coupling $g N \to \infty$. 
Based on these observations,
Gubser and Hashimoto suggested tentatively that 
the Lagrangian
\be
\label{lagrangian}
{\cal L} = {\cal L}_{SYM} + c \, (\alpha')^2 {\cal O}^{BI}_8 
\ee
might be used by itself at large $ g N$ to 
reproduce all correction coefficients
$b^l_{n,k}$ in the full absorption cross section. Here, the
coefficient $c$ is included since the normalization of ${\cal O}_8$ in
the strong coupling Lagrangian is not automatically the same as in the
Born Infeld action. An even stronger claim
in this direction has been made by Intriligator, who argued that a 
four dimensional
field theory with Lagrangian (\ref{lagrangian}) (for a particular value of
the coefficient $c$) is holographically dual to the full 
type IIB {\it string} theory on
the D3-brane geometry, for arbitrary values of $g$ and $N$.

Claims of this nature seem problematic for a variety of reasons. Firstly,
it is not clear what the theory (\ref{lagrangian}) means, since the
operator ${\cal O}_8$
is not renormalizable. One possible definition would be as a Wilsonian
action with a physical cutoff at some scale of  order  $R$; 
however, one must
somehow introduce a cutoff in the theory while preserving all the
supersymmetry. In \cite{Intriligator}, Intriligator takes the point of
view that for each value of $c$, there is a unique theory with 16
supercharges whose Lagrangian is (\ref{lagrangian}) 
in the near infrared, 
and argues that the form of the
Lagrangian for these theories is precisely (\ref{lagrangian}) along the
entire RG flow (though it is unclear what the UV fixed point could be).

Even if there is a sensible way to define a field theory based on the
Lagrangian (\ref{lagrangian}), an exact duality with supergravity/string theory
on the D3-brane
geometry would be surprising, since without taking the usual near
horizon limit, the degrees of freedom on the brane do not decouple from
those of the bulk. From this point of view, it appears that any theory
dual to supergravity on the full D3-brane geometry should include both
brane and bulk degrees of freedom, as emphasized in \cite{DGKS}\footnote{See
also \cite{hashimoto} for another approach to D3-brane holography.}.

In this note, our goals will be more modest than trying to determine
a holographic dual to the full D3-brane geometry. We simply assume that in
the near infrared, and at strong 't Hooft coupling,
the worldvolume degrees of freedom on the brane are
governed by a theory of the form (\ref{lagrangian}).  
We then compute the two point function of 
the dilaton operator in this deformation of $\N=4$
SYM theory in an attempt
to reproduce the leading correction $b^l_{1,1}$. We will see that the
required computation is independent of the regularization scheme so
all field theory calculations are well defined. Our
strategy will be to fix the coefficient of ${\cal O}_8$ in
(\ref{lagrangian}) by
requiring that the leading correction for the s-wave cross section ($l=0$)
matches with the supergravity result. Having fixed this normalization, we
compute the remaining coefficients $b^l_{1,1}$ for all higher partial
waves. We find  that the result,
\be
\label{sugb}
b^l_{1,1} = -{1 \over (l+1)(l+2)(l+3)} \; ,
\ee
is in precise agreement with the supergravity calculation of
\cite{GH}. Thus, by
choosing a single coefficient in the worldvolume theory (the normalization
of ${\cal O}_8$), we are able to reproduce the entire series of
coefficients $b^l_{1,1}$ through a calculation in the worldvolume field
theory. 
 Thus,  at least in the near infrared, 
the field theory Lagrangian (\ref{lagrangian}) appears
to provide a holographic description of the  physics on the
supergravity side, 
where the geometry is slightly deformed from pure AdS. 

We stress that we use the pure AdS/CFT correspondence 
as a tool to perform
the calculations in strongly coupled $\N=4$ SYM theory.
However, since all the required correlators are believed to obey
non-renormalization theorems, the computations could equivalently 
be performed in the free field theory limit by a direct perturbative
calculation. (We elaborate on this point in section 4).

In section 2, we review the supergravity derivation of the coefficients
$b^l_{1,1}$ to obtain the explicit expression (\ref{sugb}). In section 3, we
calculate the leading corrections to the two point function in the
worldvolume theory defined by (\ref{lagrangian}), and show that the results 
precisely match those from supergravity. In section 4, we offer some
concluding remarks.

\section{Supergravity calculation of the corrections to dilaton absorption}

\setcounter{equation}{0}

The coefficients $b_{1,1}^l$ have been implicitly determined
in \cite{GH} in terms of associated Mathieu functions.\footnote{They may be
determined explicitly from equation (22) in \cite{GH} using the first
correction to $\chi$ which may be found as equation (A.18) in \cite{DGKS}.} 
Here we sketch an alternative derivation, which is a straightforward extension
of the methods of \cite{GHKK}.

The wave equation for the $l$-th partial wave of a minimally
coupled scalar is
\be
\left[ \rho^{-5} \frac{d}{d \rho}\rho^5  \frac{d}{d \rho} +1 - 
\frac{l(l+4)}{\rho^2} + \frac{(\omega R)^4}{\rho^4}
\right] \phi(\rho) =0 \,.
\ee
Here  $\rho = \omega r$, where $r$ is the standard radial
coordinate that enters the harmonic function in the D3 brane metric,
$H(r) = 1 + R^4/r^4$. The wave equation is self-dual under the
change of variables $y = (\omega R)^2/\rho$, $\phi = y^4 \psi$,
\be
\left[ y^{-5} \frac{d}{d y}y^5  \frac{d}{d y} +1 - 
\frac{l(l+4)}{y^2} + \frac{(\omega R)^4}{y^4}
\right] \psi (y) =0 \,.
\ee
Following \cite{GHKK}, we can find 
solutions  perturbatively
in $(\omega R)^4$, both  in the ``inner'' region {\bf I} ($\rho \ll 1$)
and in the ``outer'' region {\bf III} ($\rho \gg (\omega R)^2$). To order $(\omega R)^4$,
\bea
\phi^{\bf I} &=& y^2 H_{2+l}^{(1)}(y) + 
 (\omega R)^4 \frac{\pi y^2}{2}
\int^y \frac{dx}{x^3} H^{(1)}_{2+l}(x) (J_{2+l}(x) N_{2+l}(y)-N_{2+l}(x)
J_{2+l}(y))  
\\
 \frac{\phi^{\bf III}}{A}& =& \frac{J_{2+l}(\rho)}{\rho^2} -
(\omega R)^4 \frac{\pi}
{2 \rho^2} \left[ \int^\rho \frac{d \sigma}{\sigma^3} J_{2+l}(\sigma)^2
N_{2+l}(\rho) -   \int^\rho \frac{d \sigma}{\sigma^3}J_{2+l}(\sigma) 
N_{2+l}(\sigma) J_{2+l}(\rho)  \right] \,.\nonumber
\eea
In the inner region, the solution has been chosen to satisfy
the boundary condition of purely ingoing flux at the horizon.
In the outer region, the relative ratio of the two independent
solutions of the second order wave equation 
has been fixed by the condition  that in the transition region 
$\phi^{\bf III}$ can be matched to $\phi^{\bf I}$  by a choice 
of the overall factor $A$.
 The coefficient $A$ is then be determined by requiring exact matching 
up to order 
$(\omega R)^4 \log(\omega R)$,
\be
A = \frac{-i \, 4^{2+l} (l+1)!^{\,2} \,(l+2)}{\pi (\omega R)^{2l}}\left[ 1 + 
\frac{(\omega R)^4 \log(\omega R) }{2(l+1)(l+2)(l+3)} + O((\omega R)^4)\right]\,.
\ee
To  extract the leading correction to the low-energy absorption
cross section, we simply recall that $\sigma \sim 1/|A|^2$, so that
\be
\sigma^l = \sigma_0^l \left(1-\frac{ (\omega R)^4 
\log(\omega R)}{(l+1)(l+2)(l+3)}  + O((\omega R)^4 \right) \,.
\ee
Comparing with (\ref{exact}), we see that $b^l_{1,1}$ are given by 
(\ref{sugb}).

\section{Computation in perturbed SYM theory}
\setcounter{equation}{0}

In general, the absorption cross section for a canonically normalized
field $\phi$ of frequency $\omega$ coupled to the brane by an interaction
\be
\int d^4x \; \phi(x,0) {\cal O}(x)
\ee
is given by 
\be
\sigma(\omega) = {1 \over 2i \omega} {\rm Disc} \;  \Pi(p) |_{-p^2 = \omega^2 
-i\epsilon}^{-p^2=\omega^2+i\epsilon}
\label{cross}
\ee
where $\Pi(p)$ is the momentum space two-point function,
\be
\Pi(p) = \int d^4x \; e^{i p \cdot x} \;  \langle {\cal O}(x) {\cal O}(0)
\rangle 
\ee
computed in the worldvolume theory.

In our case, $\phi$ is the $l$-th partial wave of the dilaton field and we
assume that the (strongly coupled) worldvolume theory is described by an
action\footnote{We note here that higher dimension operators, if present
in the complete Lagrangian, would not contribute to the leading correction
that we calculate.}
\be
\label{lag2}
S = \int d^4x \; ({\cal L}_{SYM} + (\alpha')^2 {\cal O}_8 )
\ee
where we have absorbed the coefficient $c$ into the definition of ${\cal
O}_8$ such that ${\cal O}_8 = c\,{\cal O}^{BI}_8$. The two point
function of the operator coupling to the $l$-th
partial wave of the dilaton in this theory is given by
\beas
\langle{\cal O}^l_\phi(x) {\cal O}^l_\phi(0)\rangle &=& 
\langle{\cal O}^l_\phi(x) \; e^{-\int
d^4 y  (\alpha')^2 {\cal O}_8(y)} \; {\cal O}^l_\phi(0)\rangle_{SYM}\\
&=& \langle{\cal O}^l_\phi(x) {\cal O}^l_\phi(0)\rangle_{SYM} 
-  (\alpha')^2 \int d^4 y \langle{\cal
O}^l_\phi(x) {\cal O}_8(y) {\cal O}^l_\phi(0)\rangle_{SYM} + \cdots
\eeas
The functional forms of these two and three point functions in $\N=4$ SYM
theory are determined completely by conformal invariance, so we may write
\be
\label{twopt}
\langle {\cal O}^l_\phi(x) {\cal O}^l_\phi(0)\rangle_{SYM} 
= {t_l \over |x|^{2l+8}}
\ee
and 
\be
\langle{\cal O}^l_\phi(x) {\cal O}_8(y) {\cal O}^l_\phi(0)\rangle_{SYM} = {r_l
\over |x|^{2l} |y|^8 |x-y|^8} \; .
\label{threept}
\ee
Using (\ref{cross}), the cross section for the $l$-th partial wave of
the dilaton field is therefore given by
\bea
\sigma = {1 \over 2i \omega} {\rm Disc} \int d^4 x \; e^{ip \cdot x }
\left\{ {t_l \over |x|^{2l+8}} -  (\alpha')^2 \int d^4 y { r_l \over |x|^{2l}
|y|^8 |x-y|^8} \right\} + \dots
\eea
The Fourier transforms and discontinuities are evaluated in the appendix.
We find
\bea
\sigma^l &=& {\pi^3 t_l \over 2^{2l+4}
(l+2)!(l+3)!} \omega^{2l+3} + {5 \pi^5 r_l  (\alpha')^2 \over 4^{l+2}
(l+4)! (l+5)!}\omega^{2l+7} \log({\omega \over \Lambda}) + \cdots
\nonumber \\
&=&\sigma^l_0 \left\{ 1 + {5 \pi^2  (\alpha')^2 \over (l+3) (l+4)^2 (l+5)}
{r_l \over t_l} \omega^4 \log({\omega \over \Lambda}) + \dots \right\}
\eea
Comparing this with the formula (\ref{exact}) for the supergravity result, we
see that\footnote{Note that we use Euclidean conventions in this section,
so $r_l$ is negative.} 
\be
\label{relation}
(b^l_{1,1} R^4)_{wv}  = {5 \pi^2  (\alpha')^2  \over (l+3) (l+4)^2
(l+5)} {r_l \over t_l} 
\ee
Thus, to evaluate $b^l_{1,1}$ we must evaluate the coefficient of the three
point function (\ref{threept}). We may take an arbitrary normalization
for ${\cal O}_\phi$ since we divide by its two point function in the result
(\ref{relation}), however the result does depend on the normalization of
${\cal O}_8$ in the action (\ref{lag2}).  On the
other hand, the quantities 
\be
\label{indep}
{b^l_{1,1} \over b^0_{1,1}} = {240 \over (l+3) (l+4)^2 (l+5)} 
{r_l t_0 \over r_0 t_l}\\
\ee
are completely independent of the normalizations used for the
operators, so these may be compared directly with the supergravity result
without worrying about the normalization of ${\cal O}_8$. 

To perform this comparison, it only remains to determine $r_l/t_l$ by 
computing the correlators (\ref{twopt}) and (\ref{threept})
in any convenient normalization. Fortunately, these exact correlation
functions  were calculated in \cite{DFMMR} via the AdS/CFT
correspondence, using the fact that ${\cal O}_l$ corresponds to
the $l$-th Kaluza-Klein mode of the dilaton field, while ${\cal O}_8$
corresponds to the dilation mode of the five-sphere.
In terms of the normalization independent
quantity $(r_l t_0)/(r_0 t_l)$, the result is 
\be
{r_l t_0 \over r_0 t_l} = {(l+4)^2 (l+5) \over 40 (l+1) (l+2)} \; .
\ee
Using (\ref{indep}), the worldvolume field theory prediction for
$b^l_{1,1}/b^0_{1,1}$ is therefore
\be
\left({b^l_{1,1} \over b^0_{1,1}}\right)_{wv} = {6 \over (l+1) (l+2) (l+3)} \; ,
\ee
which precisely agrees with the supergravity result (\ref{sugb}).

Another way to state this result is that by choosing the normalization
of ${\cal O}_8$ in the strong coupling Lagrangian such that leading
correction to the s-wave absorption cross section matches the
supergravity result, the leading corrections for all other partial
waves are also correctly reproduced. 

Actually, using the supergravity result for the s-wave correction, 
$b^0_{1,1} = -{1 \over 6}$, we are able to to give the normalization
of ${\cal O}_8$ in the strong coupling Lagrangian explicitly. We note
that equation (\ref{relation}) implies
\be
 {\pi^2 \over 48}  (\alpha')^2 {r_0 \over t_0} = b^0_{1,1} R^4 
= -{1 \over 6} (4 \pi g N (\alpha')^2)
\ee
On the other hand, Liu and Tseytlin \cite{liutseytlin} have computed
\be
{r_0 \over t_0 k} = {4 \over N \sqrt{105}} 
\ee
where $k$ is defined by
\be
\label{norm}
\langle {\cal O}_8 (x) {\cal O}_8 (0) \rangle = { k^2 \over |x|^{16}} \; .
\ee
Combining these, we conclude that the normalization of ${\cal O}_8$ in
the strong coupling Lagrangian (\ref{lag2}) is specified by
(\ref{norm}) with 
\be
k^2 = { 6720 N^2 (Ng)^2 \over \pi^2 }
\ee

\section{Remarks}
\setcounter{equation}{0}

We have demonstrated that the leading corrections to the low-energy
absorption cross sections for all partial waves of the dilaton field may
be reproduced by a field theory calculation with the Lagrangian
\be
\label{lag3}
{\cal L} = {\cal L}_{SYM} + (\alpha')^2 {\cal O}_8 
\ee
where ${\cal O}_8$ is the unique Lorentz and $SO(6)$ scalar dimension eight
operator in a short multiplet of the $\N=4$ superconformal algebra, 
normalized so that
\be
\langle {\cal O}_8 (x) {\cal O}_8 (0) \rangle =   { 6720 N^2 (Ng)^2
 \over \pi^2 }{1 \over |x|^{16}} \; .
\ee
It is interesting to note that all correlators used in our computations
(two and three point functions of chiral operators of $\N=4 $ SYM) 
are believed to obey
non-renormalization theorems, so that in principle, the calculations could
all have been performed in the free field theory approximation
without relying on
the AdS/CFT correspondence. In practice, the computation of the required
three point functions for all but the s-wave case would require knowledge
of the scalar and fermion terms in ${\cal O}_8$, and these have not yet
been determined explicitly. 

Several methods have been proposed
to obtain the precise field theory expressions of operators
dual to supergravity modes\footnote{We 
mention: correspondence
with M(atrix) theory results \cite{markmultiple, mark}; 
use of the representation theory of the $\N =4$ superconformal algebra
and superspace techniques
\cite{grouptheory}; expansion
of the DBI action in AdS background \cite{dastrivedi}; 
expansion of the flat space
DBI action in terms of the  modes determined by
the ``matching'' procedure in the supergravity absorption calculation \cite
{markmultiple}.}. We would like to 
point out that in the specific case of ${\cal O}_8$,
much of the relevant information is already available from studies
of supersymmetry in the low-energy open string theory effective action
\cite{effectiveaction, metsaev, sezgin, metsaevtseytlin}. 
Metsaev and Rahmanov \cite{metsaev}
considered ten dimensional $U(1)$ SYM theory perturbed by the most
general combination of irrelevant operators up to dimension
eight, and showed that supersymmetry restricts the form
of the action to be
\be
S_{10d} = \int d^{10}x\;  {\cal L}_{SYM     } + a\, \alpha'^2 \,
{\cal O}^{10d}_8 + O(\alpha'^3) \label{10d}
\ee
where ${\cal O}_8^{10d}$ is completely  determined 
by supersymmetry and explicitly given
in terms of ten dimensional fields, including fermions\footnote{The action
(\ref{10d}) is  invariant
under a modified supersymmetry 
$\delta=\delta_0 + \alpha'^2 \delta_1 + \dots$, where $\delta_0$ is the usual
supersymmetry of the unperturbed SYM theory. 
A simple argument shows then that
$\delta_0 \, {\cal O}_8^{10d} =0$ (up to total derivatives), 
when the SYM equations of motions
are imposed. This identifies  ${\cal O}_8^{10d}$
as the top member of a supermultiplet.}.
For a specific
value of $a$ this action correctly reproduces
the string amplitudes to order $O(\alpha'^2)$.
The operator ${\cal O}_8^{BI}$
that appears in the four dimensional abelian Born-Infeld action is (by definition) 
the dimensional reduction of ${\cal O}^{10d}_8$. The non-abelian $SU(N)$ color
structure is then introduced by the symmetrized trace prescription, which is
known to yield the correct string amplitudes to this
order in $\alpha'$ \cite{effectiveaction}.
However there is an important
caveat: this method (as any other method
that uses an integrated action) determines
${\cal O}_8$ only up to total derivative terms. This is immaterial
for the contribution of the irrelevant perturbation
to the dilaton two point functions considered
in this paper, but is important if one is interested in the
value of the two point function $\la {\cal O}_8(x)     
 {\cal O}_8(y) \ra$. The total derivative terms
can be fixed by requiring that ${\cal O}_8$ is 
a primary field of the four dimensional 
bosonic conformal group, i.e. is annihilated by
the  special conformal generators $K_\mu$.  Only
such an operator would lead to the usual conformal 
space-time dependence of two and three point functions. 
It is worth pointing out that even the standard expression
$F^4 - 1/4 (F^2)^2$ 
for the bosonic part of ${\cal O}_8$ (here $F$ denotes
the ten dimensional field strength)
does not have this property, and should be corrected 
by adding the appropriate total derivatives required by
conformal invariance.

For the s-wave case, the required three point correlation function may be
calculated perturbatively using only the known gauge field parts of ${\cal
O}_8$, however one must independently determine the correct normalization
for ${\cal O}_8$ in the Lagrangian to make a non-trivial prediction. In
fact, such a calculation was attempted in \cite{GHKK} assuming that the correct
normalization for ${\cal O}_8$ could be taken from the weak coupling
Born-Infeld Lagrangian\footnote{A similar
attempt, but using the DBI action in AdS background, was undertaken
in \cite{liutseytlin}.}. 
Their result for $b^0_{1,1}$ from the field theory
calculation was $3/4$ of the supergravity result, however the discrepancy
was partly due to an incorrect treatment of the non-abelian color
structure of
${\cal O}_8$. Using the correct  prescription, it turns out
that the field theory result is exactly $1/2$ of the supergravity result.
Since the correlators themselves are not renormalized, the most likely
explanation for this discrepancy is that the normalization of ${\cal O}_8$ in the
weak coupling Lagrangian is exactly $1/2$ of its normalization in the
strong coupling Lagrangian. If all this is correct, we are able to make a
prediction that
\be
\langle {\cal O}^{BI}_8 (x) {\cal O}^{BI}_8 (0) \rangle = {1680 N^2
(Ng)^2 \over \pi^2}{ 1 \over |x|^{16}} \; ,
\ee
where ${\cal O}^{BI}_8$ is the dimension 8 term in the Born Infeld
Lagrangian,  given in (\ref{BIop}), with the 
total derivative ambiguity in  ${\cal O}^{BI}_8$
fixed by the requirement
of conformal invariance, as explained above.

A natural extension of the work presented here would be to try to
reproduce further correction terms in (\ref{exact}) using the
Lagrangian (\ref{lag3}). It seems plausible that the higher order
correction terms of the form $b^l_{n,n}$ are independent of
regularization, so that in principle they could be 
derived in a worldvolume  calculation and directly
compared with the supergravity results. 
 If a field theory calculation using only the action (\ref{lag3}) matched with
the supergravity prediction (even for $b^0_{2,2}$), it would be even stronger 
evidence for the Gubser-Hashimoto  proposal, since generically, one
would expect possible higher dimension operators in the action
to contribute here.

On the other hand, as mentioned in the appendix, the terms $b_{1,0}$ come 
from subleading, regulation dependent 
terms in the discontinuities of the correlators.   
We do not have an unambiguous way to
evaluate them in order to compare with the supergravity result.
The problem of the proper UV definition of the action (\ref{lag3})
remains a fundamental one. It 
would be very interesting to see whether
by demanding that the $b^l_{1,0}$ terms (and other
regularization dependent terms) are correctly reproduced, one is
led to a natural renormalization scheme for  (\ref{lag3}) that would
make it well defined as a Wilsonian action.

\section*{Acknowledgements}

We would like to thank Dan Freedman, 
Steve Gubser, Ken Intriligator, Igor Klebanov, Kostas Skenderis, and
Wati Taylor  
for helpful discussions. The work of L.R. is supported in part
 by  Princeton University ``Dicke Fellowship'' and by NSF
grant 9802484.
The work of M.V.R is supported in part by the Stanford Institute 
for Theoretical Physics and by NSF grant 9870115 .  

\appendix

\section{Evaluation of Fourier transforms and discontinuities.}

\setcounter{equation}{0}

In this appendix, we determine the discontinuity across the positive
real axis in the functions
\be
f_l(p^2) = \int d^4 x \; e^{ip \cdot x} { 1 \over |x|^{2l+8}}
\ee
and
\be
g_l(p^2) = \int d^4 x \; e^{ip \cdot x} \int d^4 y {1 \over |x|^{2l}
|x-y|^8 |y|^8 }\,.
\ee
To do this, we make use of equation (41) in \cite{GH}, 
\be
\label{discs}
{\rm Disc} \int d^4x {e^{ip \cdot x} (\mu x)^{2a} \over x^{2n}} = \left( 
{\omega^2 \over 4} \right)^{n-2} \left( {4 \mu^2 \over \omega^2} \right)^a
{2 \pi^3 i \over \Gamma(n-a) \Gamma(n-a-1) } \; . 
\ee
Using this relation for $a=0$, we find immediately that
\be
{\rm Disc} \; f_l(p^2) |_{p^2 = \omega^2 - i \epsilon}^{p^2 = \omega^2  +
i \epsilon} = 2 \pi i { \pi^2 \omega^{2l + 4} \over 4^{l+2} (l+2)! (l+3)! }\,.
\ee
To evaluate $g_l(p^2)$, we first perform the integral over $y$. This
requires regularization, which we implement
with the cutoff $|y|,|y-x| > {1 \over
\Lambda}$. The resulting integral has leading terms with various positive
powers of $\Lambda$. 
 As standard in field theory,
these divergent contributions could be reabsorbed with local counterterms.
This requires
a renormalization prescription, and we do not
have an a priori natural choice. However,
the correction to the cross section that we are 
interested in
(proportional to $\omega^4 \log(\omega)$) 
arises from the term logarithmic
in $x^2$, whose coefficient is independent of the cutoff. Explicit computation
gives
\be
{40 \pi^2 \over x^{2l+12}} \log(x^2 \Lambda^2) \,.
\ee
The discontinuity in the Fourier transform of this function may be read off 
from the order $a$ term in the expansion of equation (\ref{discs}), and we 
find   
\be
{\rm Disc} \; g_l(p^2) |_{p^2 = \omega^2 - i \epsilon}^{p^2 = \omega^2  +
i \epsilon} =  - 2 \pi i {5 \pi^4 \over 2^{2l+5} (l+4)! (l+5)!}
\omega^{2l+8} \log \left( \omega^2 \over \Lambda^2 \right) + \cdots \; .
\ee
The dots denote a single additional term proportional to $p^{2l+8}$
without a logarithm. This term is responsible for the $b^l_{1,0}$ term
in the cross section, however it is dependent on the
renormalization prescription. 
In contrast, the coefficient of the leading term in the
discontinuity is universal. A way to understand the universality
of the logarithmic coefficient is its relation with the
conformal anomaly of the field theory 
in the presence of external sources for composite operators,
 see e.g. \cite{kostas}\footnote{We thank K. Skenderis for discussions on this
 point.}.

\end{document}